# Resultant Delta-V Estimation from EDR Data Recorded in Automobiles that have Undergone Impact-Induced Yaw Rate

Micky Marine
SSi Phoenix, Inc.

There are several references in the public literature that discuss the effect impact-induced yaw motion has on the measurement of acceleration, vis-à-vis accelerometers, in automobile collisions [1], [2], [3], [4]. It is well-understood that direct integration of accelerometer data does not provide accurate velocity components for a vehicle undergoing appreciable rotational motion whether the accelerometers are installed at the vehicle center of gravity or not. Direct integration of accelerometer data is, nonetheless, how event data recorders (EDRs) calculate Delta-V components and care must be taken on the part of the analyst in interpreting this information when the vehicle from which it came was known to have experienced significant yaw motion. As such, in this paper we set out to: (1) examine whether the correct resultant Delta-V at the center of gravity can be determined from the directly-integrated EDR Delta-V components, and (2) to assess what useful Delta-V information can be readily gotten from EDRs that are typically not installed at the vehicle center of gravity. Before digging into these tasks, however, a summary of the relevant equations is in order.

Referring to Figure 1, the absolute acceleration and velocity vectors can be expressed as stated below in equations (1) through (4). While these equations apply to any point on the vehicle, for now we'll consider the vectors at the vehicle center of gravity (CG). Also included equations (1) through (4) are the vector transformations to and from the ground-based coordinate system to the vehicle-based coordinate system.

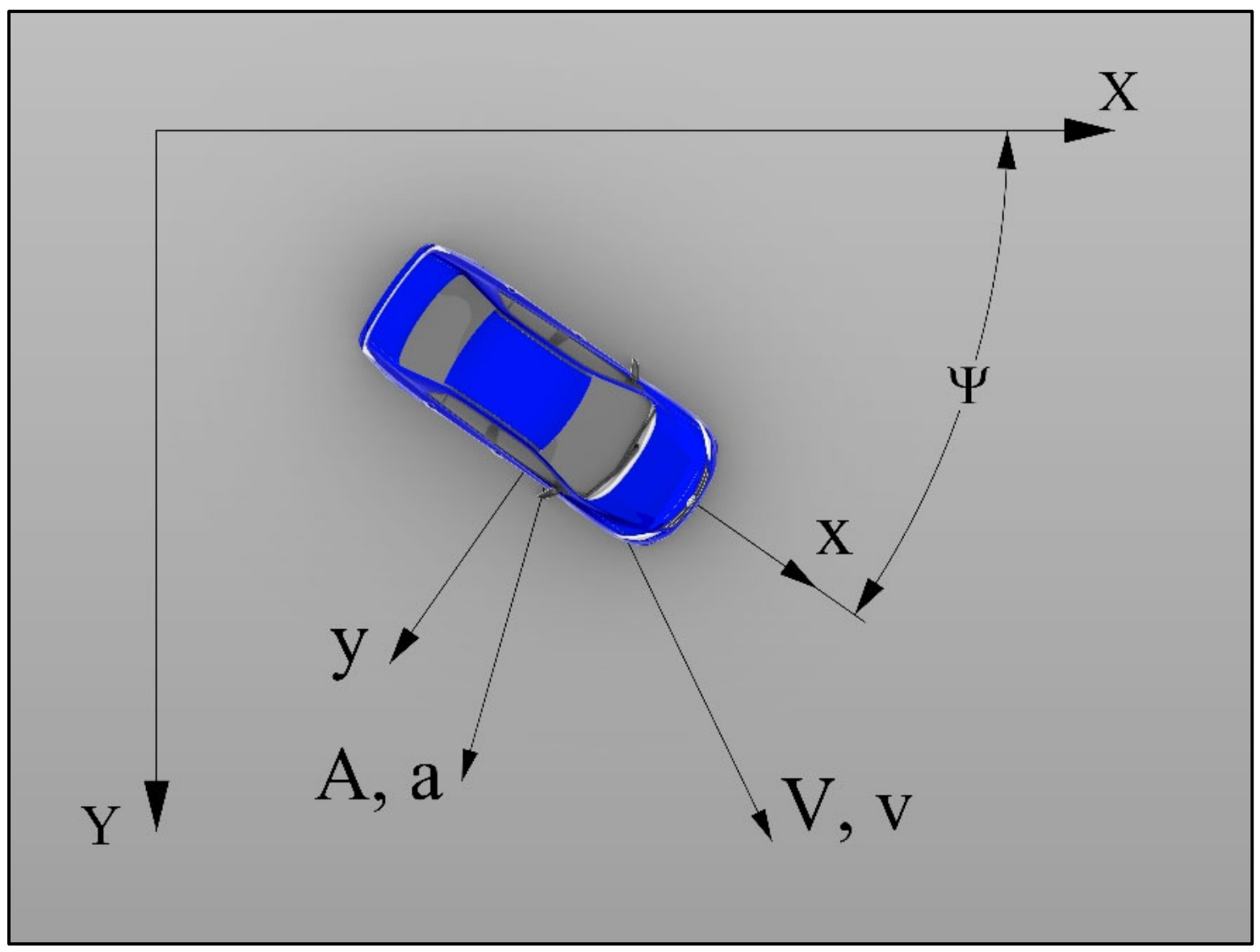

Figure 1: Acceleration and velocity vectors within ground-based and vehicle-based reference systems

$$\vec{A} = A_x\hat{\imath} + A_y\hat{\jmath} \qquad \vec{a} = a_x\hat{e}_1 + a_y\hat{e}_2$$

$$\begin{bmatrix} a_x \\ a_y \end{bmatrix} = \begin{bmatrix} \cos\psi & \sin\psi \\ -\sin\psi & \cos\psi \end{bmatrix} \begin{bmatrix} A_x \\ A_y \end{bmatrix} = \underline{T}\vec{A} \tag{1}$$

$$\begin{bmatrix} A_x \\ A_y \end{bmatrix} = \begin{bmatrix} \cos\psi & -\sin\psi \\ \sin\psi & \cos\psi \end{bmatrix} \begin{bmatrix} a_x \\ a_y \end{bmatrix} = \underline{T}^{-1}\vec{a} \tag{2}$$

$$\vec{V} = V_x\hat{\imath} + V_y\hat{\jmath} \qquad \vec{v} = v_x\hat{e}_1 + v_y\hat{e}_2$$

$$\begin{bmatrix} v_x \\ v_y \end{bmatrix} = \begin{bmatrix} \cos\psi & \sin\psi \\ -\sin\psi & \cos\psi \end{bmatrix} \begin{bmatrix} V_x \\ V_y \end{bmatrix} = \underline{T}\vec{V} \tag{3}$$

$$\begin{bmatrix} V_x \\ V_y \end{bmatrix} = \begin{bmatrix} \cos\psi & -\sin\psi \\ \sin\psi & \cos\psi \end{bmatrix} \begin{bmatrix} u \\ v \end{bmatrix} = \underline{T}^{-1}\vec{v} \tag{4}$$

The absolute acceleration vector can be derived through the differentiation of the absolute velocity vector. In the ground-based system, this is expressed as:

$$\vec{A} = \frac{d\vec{V}}{dt} = \frac{d(V_x\hat{\imath})}{dt} + \frac{d(V_y\hat{\jmath})}{dt} = \left(\frac{dV_x}{dt}\hat{\imath} + V_x\frac{d\hat{\imath}}{dt}\right) + \left(\frac{dV_y}{dt}\hat{\jmath} + V_y\frac{d\hat{\jmath}}{dt}\right)$$

$$\vec{A} = \frac{d\vec{V}}{dt} = \frac{dV_x}{dt}\hat{\imath} + \frac{dV_y}{dt}\hat{\jmath} = \dot{V}_x\hat{\imath} + \dot{V}_y\hat{\jmath} \tag{5}$$

In the ground-based (fixed) reference system we get a very straightforward result because the unit vectors that define that system are, by definition, not changing orientation. As we see in the equations below, this is not the case for differentiation of the velocity vector expressed in the vehicle-based reference system, where the unit vectors defining that system can change orientation.

$$\vec{a} = \frac{d\vec{v}}{dt} = \frac{d(v_x\hat{e}_1)}{dt} + \frac{d(v_y\hat{e}_2)}{dt} = \left(\frac{dv_x}{dt}\hat{e}_1 + v_x\frac{d\hat{e}_1}{dt}\right) + \left(\frac{dv_y}{dt}\hat{e}_2 + v_y\frac{d\hat{e}_2}{dt}\right)$$

$$\vec{a} = \frac{d\vec{v}}{dt} = (\dot{v}_x\hat{e}_1 + \vec{\omega} \times v_x\hat{e}_1) + (\dot{v}_y\hat{e}_2 + \vec{\omega} \times v_y\hat{e}_2); \quad where\ \frac{d\hat{e}_i}{dt} = \vec{\omega} \times \hat{e}_i$$

$$\vec{a} = \frac{d\vec{v}}{dt} = (\dot{v}_x\hat{e}_1 + v_x\omega\hat{e}_2) + (\dot{v}_y\hat{e}_2 - v_y\omega\hat{e}_1)$$

$$\vec{a} = \frac{d\vec{v}}{dt} = (\dot{v}_x - v_y\omega)\hat{e}_1 + (\dot{v}_y + v_x\omega)\hat{e}_2 \tag{6}$$

Note in this sequence of calculations that there is a regrouping in the final step to get to equation (6) in which differentiation contributions from both the longitudinal and lateral velocity vector components are collected to form the two respective vehicle-based acceleration components. The ramification of this regrouping is that to directly integrate the vehicle-based acceleration vector components of equation (6) is to apply the incorrect anti-derivatives with which to return to a valid velocity vector. While the above vehicle-based acceleration components are valid physical measures, the direct integration of the components do not produce useful velocity or Delta-V information and no physical meaning can be assigned to these integration operations when based on accelerometer data from a vehicle undergoing significant yaw motion.

To obtain vehicle Delta-V from accelerometer data, however, an integration process must occur. In the ground-based system, this is accomplished very straightforwardly.

$$\Delta\vec{V} = \int \frac{dV_x}{dt} dt\,\hat{\imath} + \int \frac{dV_y}{dt} dt\,\hat{\jmath} = (V_x - V_{x0})i + (V_y - V_{y0})\hat{\jmath} = \Delta V_x \hat{\imath} + \Delta V_y \hat{\jmath} \tag{7}$$

Once again, things are not so simple when integrating in the vehicle-based system. If we rearrange the components of equation (6), and have sufficient yaw rate information, we can integrate, as shown in equations (8) and (9), to obtain the vehicle-based velocity vector components:

$$\int \dot{v}_x dt = v_x - v_{x0} = \int (a_x + v_y \omega) dt \tag{8}$$

$$\int \dot{v}_y dt = v_y - v_{y0} = \int (a_y - v_x \omega) dt \tag{9}$$

The integrations of equations (8) and (9) provide us the velocity vector component time-histories in the vehicle-reference system but, as discussed in reference [2], this does not get us to a proper accounting of the absolute Delta-V expressed in this reference frame. The correct Delta-V components expressed in the vehicle-based reference system are not merely expressed as,

$$\Delta v_x \neq v_x - v_{x0}$$

$$\Delta v_y \neq v_y - v_{y0}$$

To correctly calculate Delta-V at a particular time using the vehicle-frame velocity components, the initial velocity vector must be transformed to the current vehicle reference frame orientation.

$$\Delta\vec{v}_{CG} = \underline{T}\Delta\vec{V}_{CG} = \underline{T}(\vec{V}_{CG} - \vec{V}_{CG0}) = \underline{T}\left(\vec{V}_{CG} - \underline{T_0}^{-1}\vec{v}_{CG0}\right) = \begin{bmatrix} v_x \\ v_y \end{bmatrix} - \underline{T}\,\underline{T_0}^{-1}\begin{bmatrix} v_{x0} \\ v_{y0} \end{bmatrix}$$

$$\Delta\vec{v}_{CG} = \begin{bmatrix} v_x \\ v_y \end{bmatrix} - \underline{T}' \begin{bmatrix} v_{x0} \\ v_{y0} \end{bmatrix}; \quad \underline{T}' = \underline{T}\,\underline{T_0}^{-1} = \begin{bmatrix} \cos(\psi - \psi_0) & \sin(\psi - \psi_0) \\ -\sin(\psi - \psi_0) & \cos(\psi - \psi_0) \end{bmatrix} \tag{10}$$

Given the Delta-V components so far defined, we can clearly state that the following expressions for the absolute resultant Delta-V vector are all correct and equal to one another:

$$|\Delta\vec{V}| = \sqrt{\left[\int A_x dt\right]^2 + \left[\int A_y dt\right]^2} = \sqrt{\Delta V_x^2 + \Delta V_y^2} = \sqrt{\Delta v_x^2 + \Delta v_y^2}$$

Can it be said, however, that the resultant of the integrals of the acceleration components measured in the rotating vehicle-based reference frame is the same as the resultant of the absolute Delta-V vector? Stated mathematically, is the following expression in fact true?

$$\sqrt{\left[\int a_x dt\right]^2 + \left[\int a_y dt\right]^2} =? \sqrt{\left[\int A_x dt\right]^2 + \left[\int A_y dt\right]^2} \tag{11}$$

The acceleration components within equation (11) are all valid measures of the absolute acceleration in their respective reference frames. Prior discussions regarding yaw effects on accelerometer data have stated that calculating the resultant of the directly integrated vehicle-referenced acceleration data "should … provide an accurate scalar, resultant Delta-V." [2], and "…if the EDR is located at the vehicle's CG, the resultant delta-v calculated by the EDR matches the resultant absolute delta-v…" [4]. To examine whether equation (11) is in fact a true mathematical expression, we rewrite the vehicle-referenced acceleration components in terms of the acceleration components referenced to a fixed system.

$$\sqrt{\left[\int\left(A_x \cos\psi + A_y \sin\psi\right)dt\right]^2 + \left[\int\left(-A_x \sin\psi + A_y \cos\psi\right)dt\right]^2} =? \sqrt{\left[\int A_x dt\right]^2 + \left[\int A_y dt\right]^2} \quad (12)$$

If we were working with the squares of the respective integrands, we would certainly be working with a true mathematical statement. Additionally, if there is no rotational change during the impulse, then we again have a true mathematical statement. When vehicle rotation occurs, and since we're working with the squares of the integrals rather than that of the integrands, matters are not so simple. To help flesh things out, we'll look at a simple example where the ground-based acceleration components are constant and equal (i.e., Ax = Ay = 1), and the angular displacement is based on a constant yaw rate (i.e., $\psi = \omega t$):

Making these substitutions and expanding the equation, we can reduce to the following expression:

$$\sqrt{2\left(\int \cos\omega t\, dt\right)^2 + 2\left(\int \sin\omega t\, dt\right)^2} =? \sqrt{2\left[\int dt\right]^2}$$

Integrating the interior terms over the interval 0 to t and expanding the squared terms, we get:

$$\sqrt{\frac{4}{\omega^2}\left(1 - \cos\omega t\right)} =? \sqrt{2t^2}$$

Using the Taylor series expansion for the cosine term we get:

$$\sqrt{\frac{4}{\omega^2}\left(1 - \left(1 - \frac{(\omega t)^2}{2!} + \frac{(\omega t)^4}{4!} - \frac{(\omega t)^6}{6!} + \cdots\right)\right)} \neq \sqrt{2t^2} \quad (13)$$

We now see that, for this simple example, our original expression of equation (11) is not true. It can be made true only if we neglect all but the two lowest-ordered terms of the Taylor series expansion. In doing so, we can get an exact match to the absolute Delta-v determined by integrating the space-fixed acceleration components:

$$\sqrt{\frac{4}{\omega^2}\left(1 - \left(1 - \frac{(\omega t)^2}{2!}\right)\right)} = \sqrt{4 \cdot \frac{t^2}{2!}} = \sqrt{2t^2}$$

Which means, of course, that calculating the resultant of the directly-integrated vehicle-based acceleration components can, at best, provide an approximation of the true absolute resultant Delta-V. Rearranging the left-hand side of equation (13) allows us to see that the use of this simple example readily distills the effects that increased time and increased yaw rate have on amplifying the difference between the resultant of the integrated vehicle-reference components and the resultant of the actual absolute Delta-V vector:

$$\delta = \sqrt{2t^2} - \sqrt{2t^2 - 4\left(\frac{\omega^2 t^4}{4!} - \frac{\omega^4 t^6}{6!} + \cdots\right)} \quad (14)$$

The error characteristics of equation (14) are shown graphically in Figure 2. In this graph we can see that the directly-integrated vehicle-based resultant provides a close approximation to the absolute Delta-V resultant for smaller time durations. As both the duration and yaw rate increase, however, the error is observed to increase.

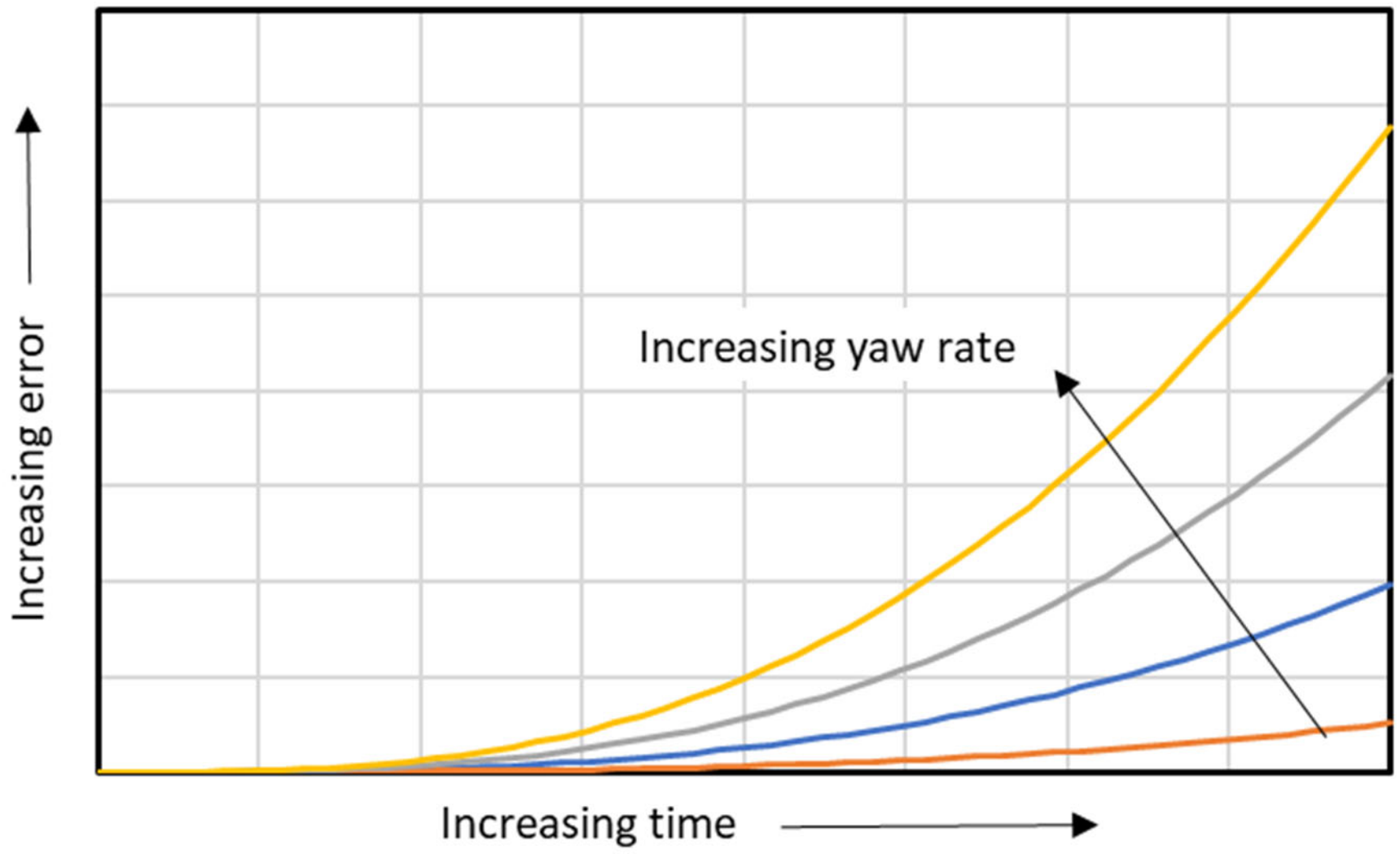

Figure 2: Error characteristics of simple example

For an example using impulsive accelerations, more akin to an automobile collision, we'll use a haversine function along the lines of which Smith [4] utilized to model a collision impulse. Figure 3 depicts the impulse configuration for our example. The relevant equations that stem from this versine function for our example are:

$$\Delta V_{YCG}(t) = \frac{\Delta V_{YCG}}{t_P}\left(t - \frac{t_P}{2\pi}\sin\frac{2\pi}{t_P}t\right) \tag{15}$$

$$\omega(t) = \frac{d}{k_z^2}\Delta V_{YCG}(t) \tag{16}$$

Where $\Delta V_{YCG}$ is the total Delta-V and is aligned with the ground-based Y-axis; $t_P$ is the impulse duration; $\omega$ is the vehicle yaw rate; d is the moment arm relative to the vehicle center of gravity that the impulse acts on the vehicle; and $k_z$ is the vehicle radius of gyration about its z-axis.

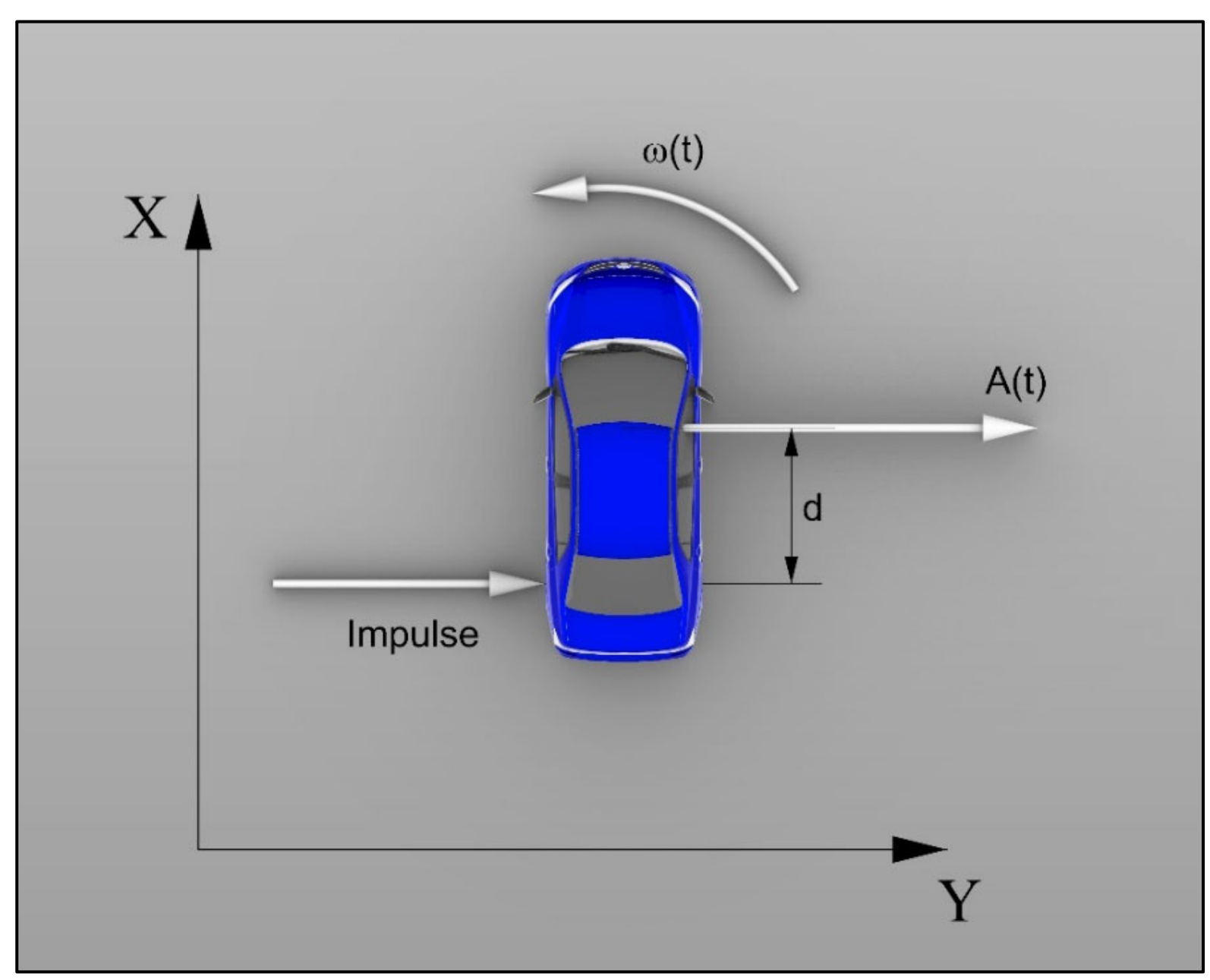

Figure 3: Example Impulse Configuration

Figure 4 depicts two plot lines in each graph: one is the analytically known absolute CG Delta-V and the other is the resultant of the directly integrated vehicle-based acceleration components, which were determined through coordinate transformation. Additionally, the impulse duration for this example was 100 milliseconds; the impulse moment arm was three feet (being aft of the vehicle CG, this is a negative value), resulting in a counter-clockwise yaw rate of around 250 degrees per second; and the radius of gyration is 4.5 feet. As we can see from this figure, the resultant of the directly-integrated vehicle-based acceleration components very closely approximates the known absolute resultant CG Delta-V.

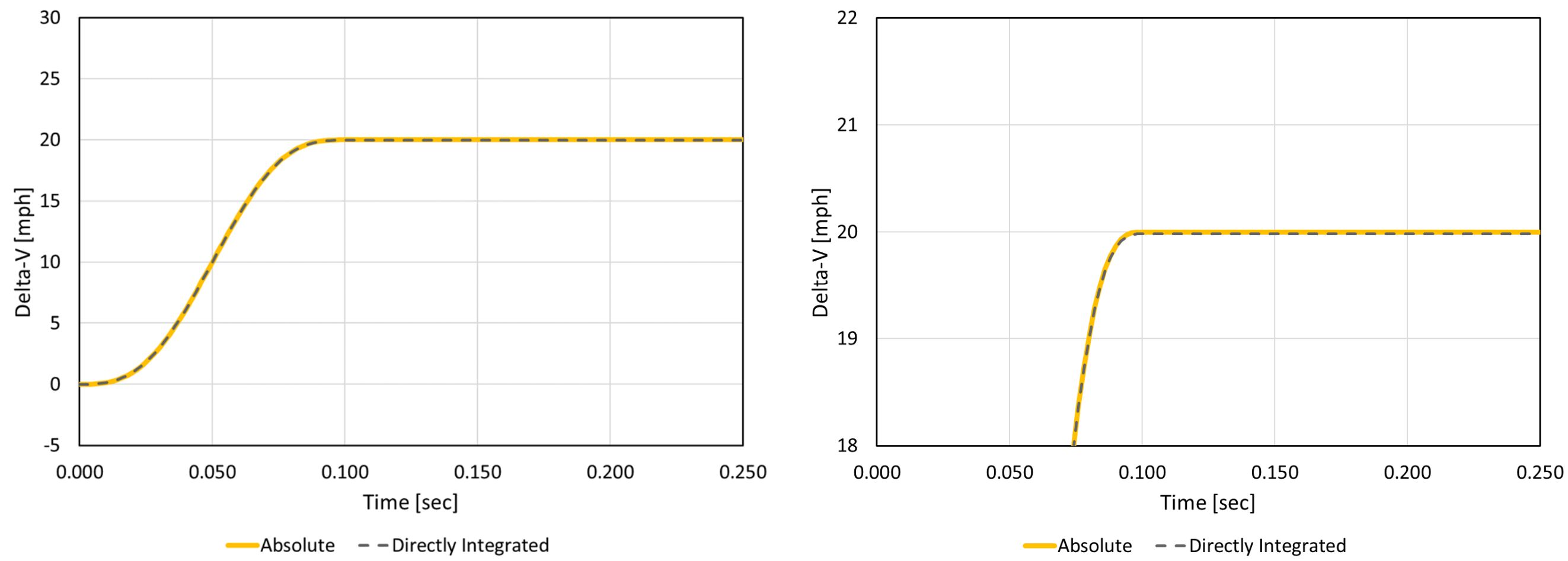

Figure 4: Resultant Delta-V: Haversine pulse – moderate yaw rate

Figure 5 depicts plots of these same values for the impact configuration in which the impulse offset distance is six feet aft of the CG. While not beyond the realm of automobile yaw-plane impact possibilities, this results in the uncommonly high yaw rate of about 500 degrees per second. In this figure we see the effects of increased yaw rate with an increased difference, though we still get a close approximation of the resultant Delta-V through the direct integration of the vehicle-based acceleration components. The effects of increased impulse duration are demonstrated in Figure 6 where we see more pronounced differences for a 150-millisecond impulse duration than those of the 100-millisecond pulse for the otherwise same conditions. Nonetheless, even for the condition of very high yaw rate and long duration impulse, the difference is a fraction of a mile per hour for our example impact configuration. Thus, while not a fundamentally proper method to determine the absolute resultant Delta-V, the resultant of the directly-integrated vehicle-based acceleration components measured or calculated at the CG can provide a simply-calculated, close approximation to the absolute resultant CG Delta-V.

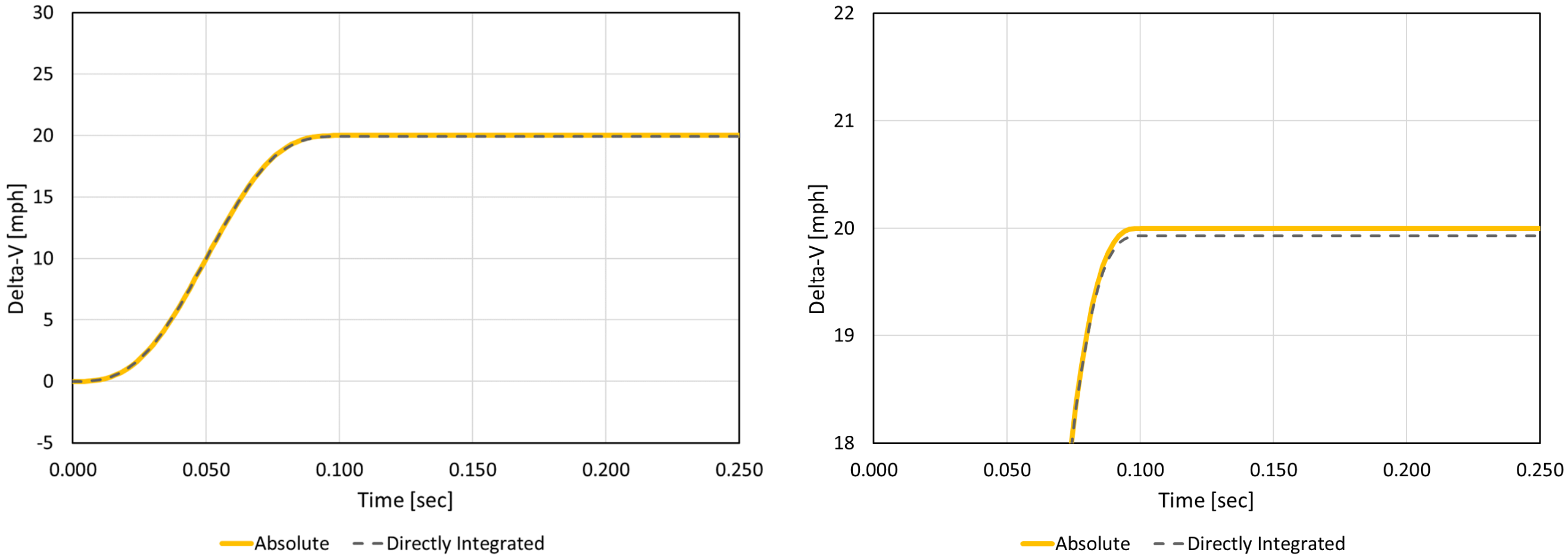

Figure 5: Resultant Delta-V: Haversine pulse – high yaw rate

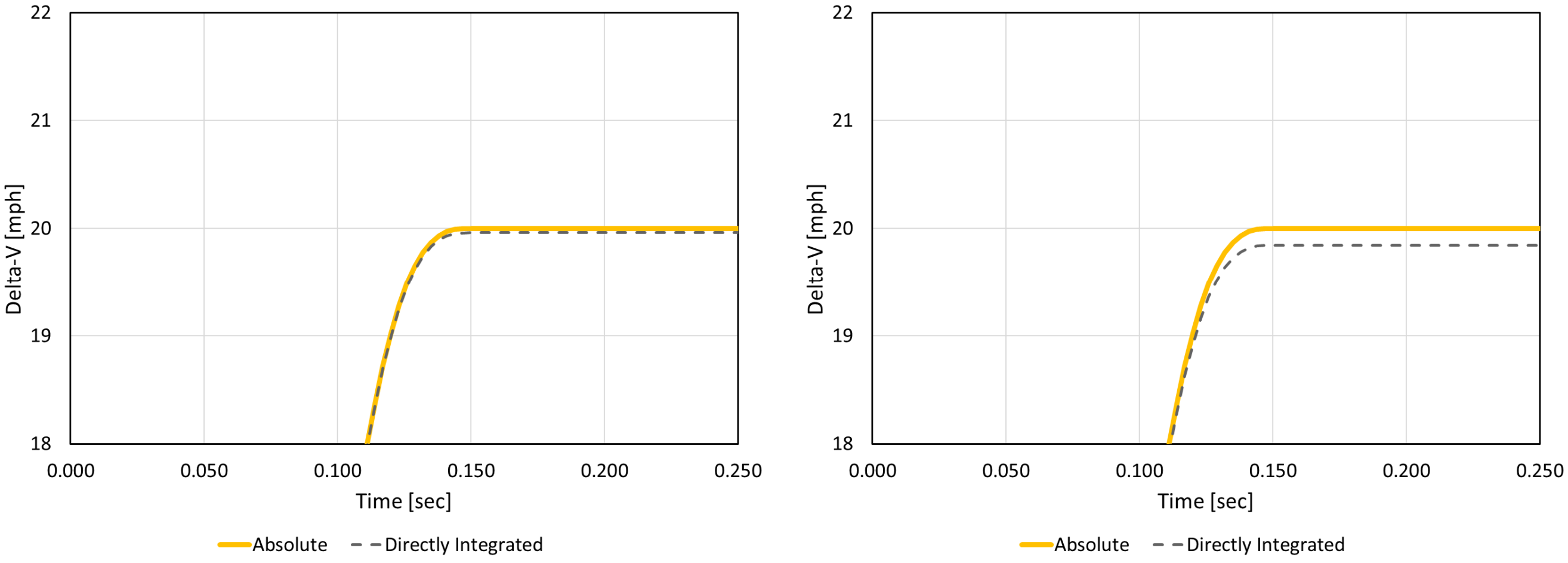

Figure 6: Resultant Delta-V with increased pulse duration: moderate yaw rate (left); increased yaw rate (right)

The above example was premised on the convenient fact that the acceleration components of the vehicle at its CG were prescribed, and the Delta-V therefore known. When working with EDR data, it is frequently not the case that CG Delta-Vs are known. In older-model vehicles, one may well find an EDR under the driver seat or the front passenger seat. In newer-model vehicles (particularly since about the 2010 model year), EDRs are generally located near the vehicle longitudinal centerline and are often located under the center stack and typically somewhat forward of the vehicle CG, while others may be positioned under the center console between the front seats and potentially aft of the CG.

An example of off-CG EDR Delta-V from a crash test with an active EDR was discussed in reference [3] and involved a 1997 Buick LeSabre traveling 60.3 mph striking the left rear of a 1998 Buick LeSabre traveling 20.5 mph. A diagram of the impact configuration is provided in Figure 7. The EDR of interest in this example was located on the target vehicle (1998 LeSabre) and measured to be 4.2 inches aft, and 19.0 inches right, of the vehicle CG. Accelerometers were mounted on the EDR and the data from these is presented in Figure 8 where the measured longitudinal and lateral accelerations are shown in the left panel and the directly-integrated Delta-V components derived from this data, along with their resultant, are shown in the right panel. Also included in the right panel of Figure 8 is the properly corrected resultant Delta-V at the CG.

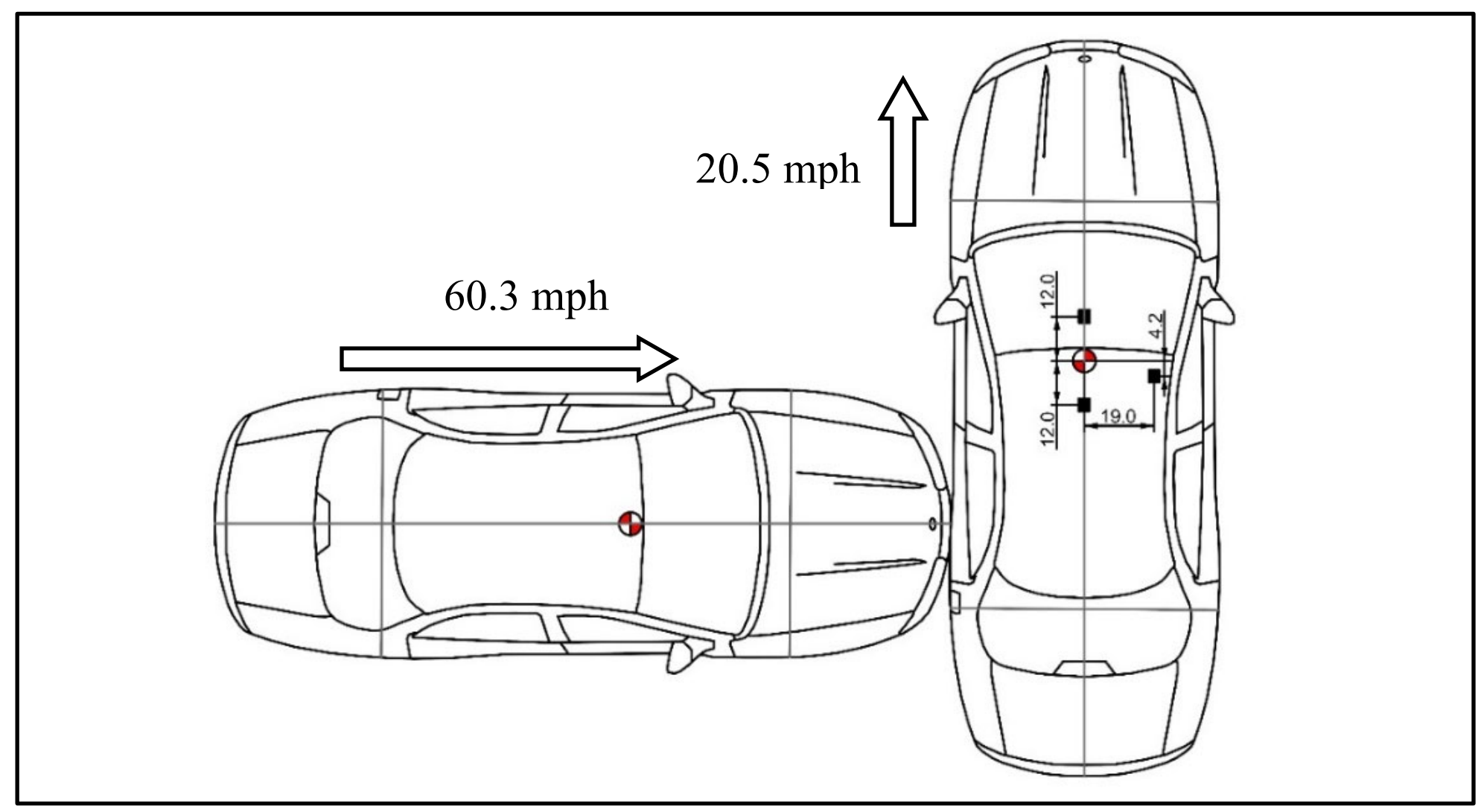

Figure 7: Plan view diagram of example crash test impact configuration

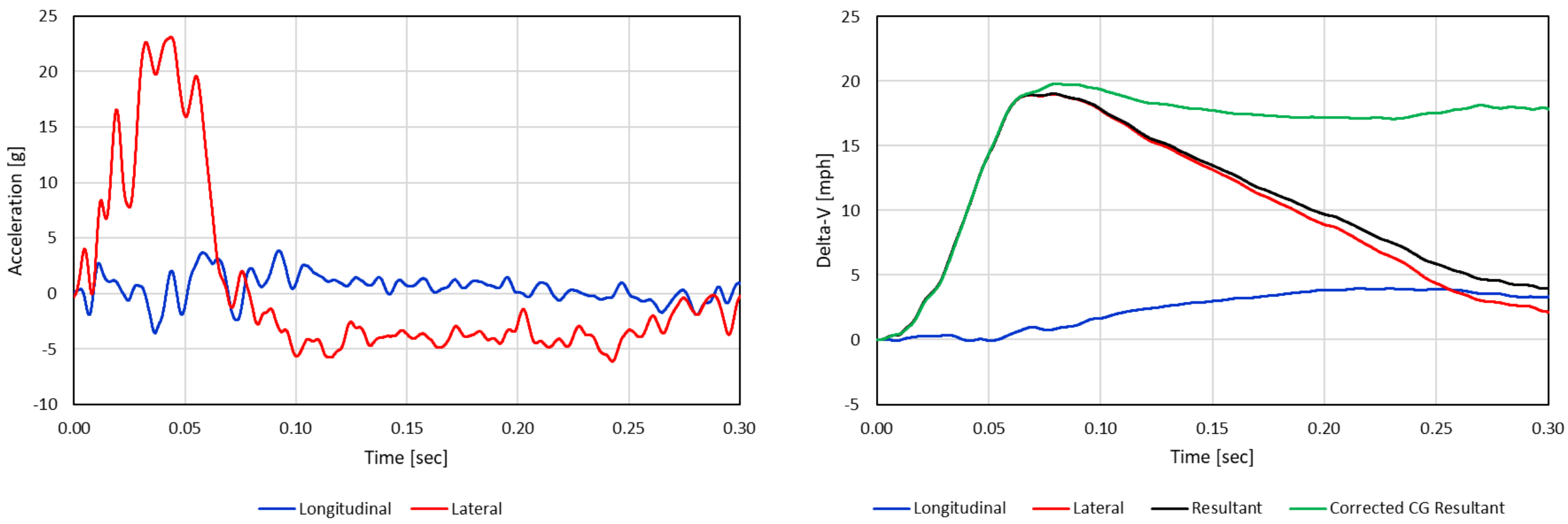

Figure 8: Acceleration and directly-integrated Delta-V from accelerometers mounted to an off-CG EDR

The EDR of the example vehicle is of a vintage that recorded only the longitudinal Delta-V, and that being only for front-to-rear changes in velocity. This is evident when comparing the directly integrated longitudinal Delta-V from the EDR-mounted accelerometers to the imaged crash pulse data plot shown in Figure 9. There are some similarities between these two sets of data with very low-level Delta-Vs recorded early on in the pulse. After about 50 milliseconds the crash test data indicates positive longitudinal Delta-V values which this EDR does not document. Since the instrumentation set of this crash test included both longitudinal and lateral accelerometers mounted to the vehicle EDR, the data from these accelerometers will be used to represent the EDR acceleration for the following analysis.

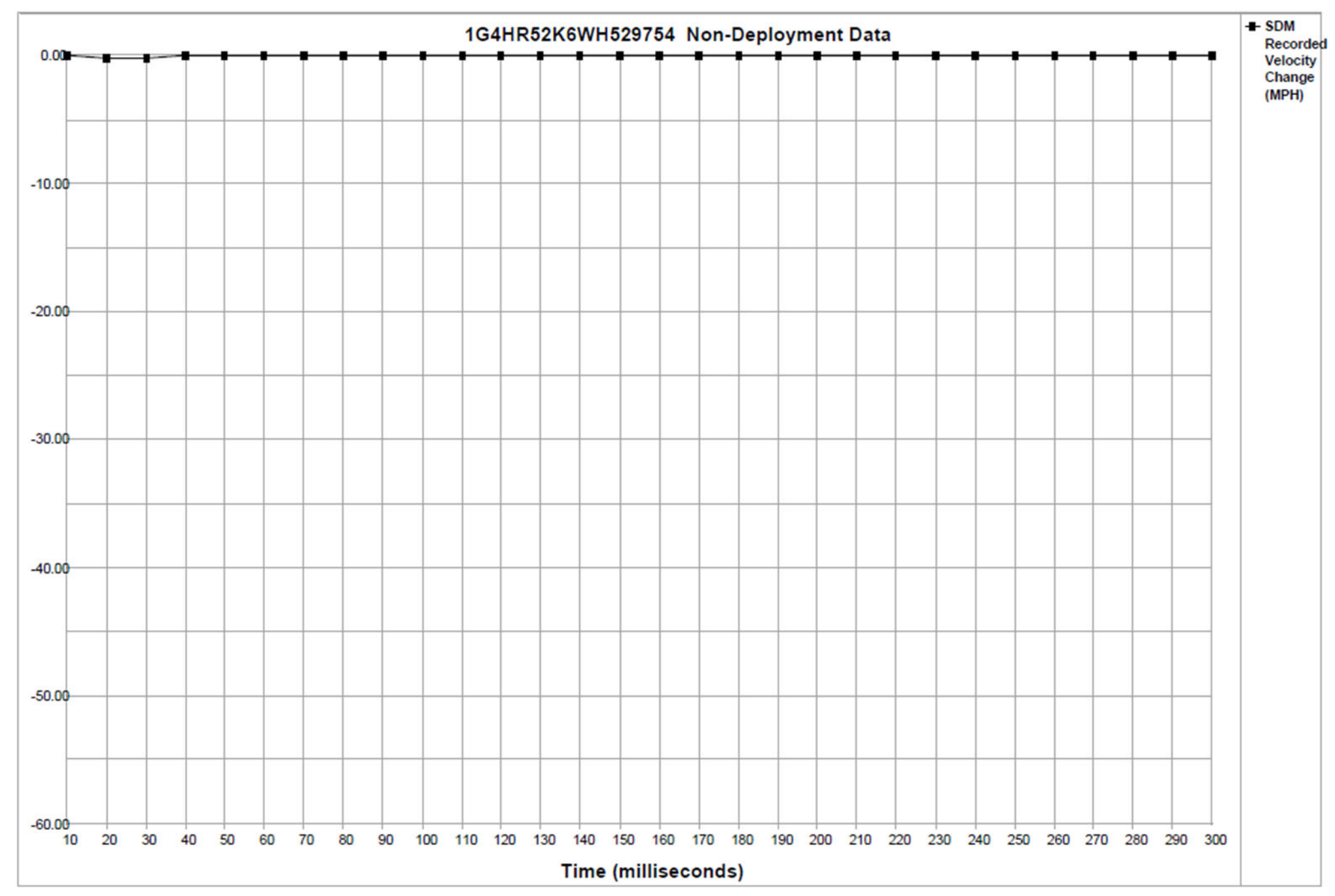

Figure 9: Longitudinal Delta-V in 1998 Buick LeSabre ACM image

One relationship between Delta-Vs at a non-CG location and those at the CG begins with the transformation of velocities from the ground-based system to the vehicle-based system, the development of the rigid-body kinematic equation relating the vehicle-based Delta-V at the CG ($\Delta v_{CG}$) to the vehicle-based Delta-V at the EDR location ($\Delta v_E$) is shown below.

$$\Delta\vec{v}_E = \underline{T}\Delta\vec{V}_E = \underline{T}\left(\vec{V}_{CG} + \vec{\omega} \times \underline{T}^{-1}\vec{r}_E\right) - \underline{T}\left(\vec{V}_{CG0} + \vec{\omega}_0 \times \underline{T_0}^{-1}\vec{r}_E\right)$$

$$\Delta\vec{v}_E = \Delta\vec{v}_{CG} + \vec{\omega} \times \vec{r}_E - \vec{\omega}_0 \times \underline{T}'\,\vec{r}_E$$

For the example crash test used here, and for many other crash tests and real-world automobile collisions, the initial yaw rate is zero (or negligibly small) and the above equation reduces to:

$$\Delta\vec{v}_{CG} = \Delta\vec{v}_E - \vec{\omega} \times \vec{r}_E \tag{17}$$

If the correct Delta-V components at the EDR location are known, along with a yaw rate time history, this equation can be used to determine the correct actual Delta-V at the CG. However, as Smith correctly noted in reference [4], because EDR-integrated Delta-V components in the presence of significant yaw rate are inaccurate, equation (17) will not result in an accurate absolute Delta-V vector time-history at the vehicle CG.

A second kinematic expression relating remote-location Delta-V to that at the CG makes use of the rigid-body relative acceleration equations that can be found in any undergraduate dynamics text book,

$$\vec{a}_E = \vec{a}_{CG} + \vec{\alpha} \times \vec{r}_E + \vec{\omega} \times \vec{\omega} \times \vec{r}_E \tag{18}$$

By integrating equation (18), along with some algebraic rearranging, we can arrive at the following equation for the directly-integrated CG acceleration vector,

$$\int \vec{a}_{CG} dt = \Delta \vec{v}_{CG}^{dir} = \int \vec{a}_E dt - \int (\vec{\alpha} \times \vec{r}_E) dt - \int (\vec{\omega} \times \vec{\omega} \times \vec{r}_E) dt \tag{19}$$

This can be further reduced to,

$$\Delta \vec{v}_{CG}^{dir} = \Delta \vec{v}_E^{dir} - \vec{\omega} \times \vec{r}_E - \int (\vec{\omega} \times \vec{\omega} \times \vec{r}_E) dt \tag{20}$$

The components of this vector equation are,

$$\Delta v_{CGx}^{dir} = \Delta v_{Ex}^{dir} + r_{Ey} \omega + \int r_{Ex} \omega^2 dt \tag{21}$$

$$\Delta v_{CGy}^{dir} = \Delta v_{Ey}^{dir} - r_{Ex} \omega + \int r_{Ey} \omega^2 dt \tag{22}$$

Equation (20) is simply equation (17) with a subtracted integrated-centripetal-acceleration term included. This tells us that, though the use of equation (17) has been recommended elsewhere as a correction for off-CG EDR locations, it is not a truly proper correction when using directly-integrated EDR data. Equation (20), however, can be used to at least produce a directly-integrated resultant CG Delta-V. As discussed above, we know that the resultant of directly integrated CG acceleration components can provide a close approximation of the actual resultant delta-V at the CG.

To calculate this resultant, however, requires a vehicle yaw rate time-history. As reported in reference [3], the peak yaw rate was measured with an angular rate gyro to be approximately 400 degrees per second, reaching this value at approximately 100 milliseconds into the crash pulse. For our current purposes, we'll work with estimates that bracket this value to examine what effect this has. A reconstructionist working a real-world accident will often not have impact-induced yaw rate data to work with and would need to arrive at some estimation of it. This might be based on available evidence that informs the analyst or a range of values the analyst deems a sufficient bracketing. In any case, the yaw rate time-history can be simply modeled with the following integrated-haversine function, where $t_p$ is the estimated impulse duration:

$$\omega(t) = \frac{\omega_P}{t_p} \left( t - \frac{1}{\lambda} \sin \lambda t \right); \quad \lambda = \frac{2\pi}{t_p} \tag{23}$$

Alternatively, if an analyst instead has information that allows for the determination of a heading angle change value (or range) at the end of the impulse ($\Delta\psi$), the integrated-haversine function below can be used as a yaw rate time-history:

$$\omega(t) = \frac{2\Delta\psi}{t_p^2} \left( t - \frac{1}{\lambda} \sin \lambda t \right); \quad \lambda = \frac{2\pi}{t_p} \tag{24}$$

For this example, we'll use a range of end-of-impulse yaw rates of 250 to 500 degrees per second, which fairly widely brackets the measured value of 400 degrees per second, and a time-to-peak yaw rate of 100 milliseconds. These yaw rate models, along with the measured yaw rate data are plotted in Figure 10.

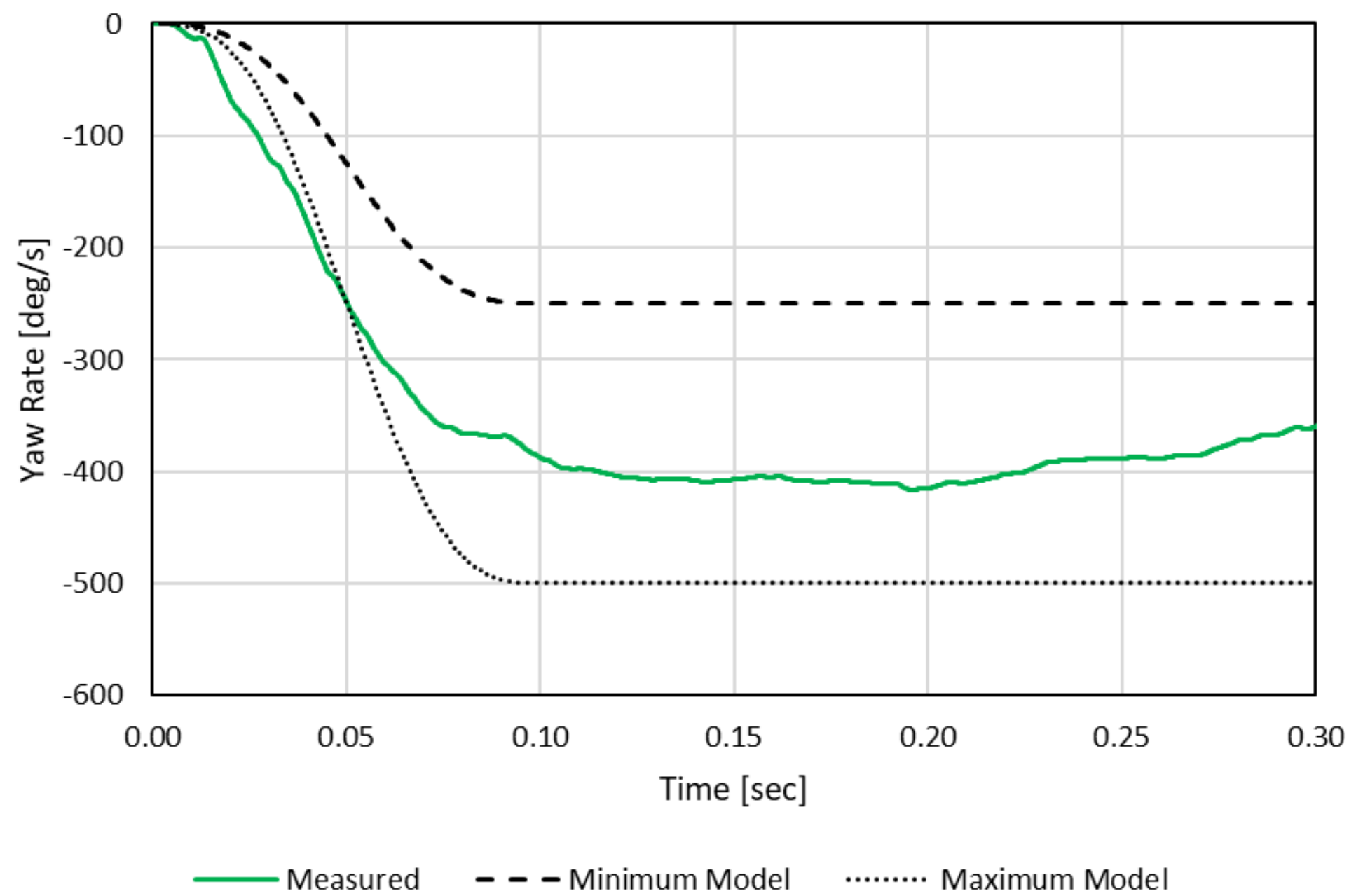

Figure 10: Measured yaw rate and minimum/maximum models

Using these yaw rate models and equations (21) and (22), the resultant CG Delta-V is plotted and shown in Figure 11. In this figure we see that the corrected CG resultant Delta-V brackets the actual peak CG resultant Delta-V by a range of about ± 2 mph.

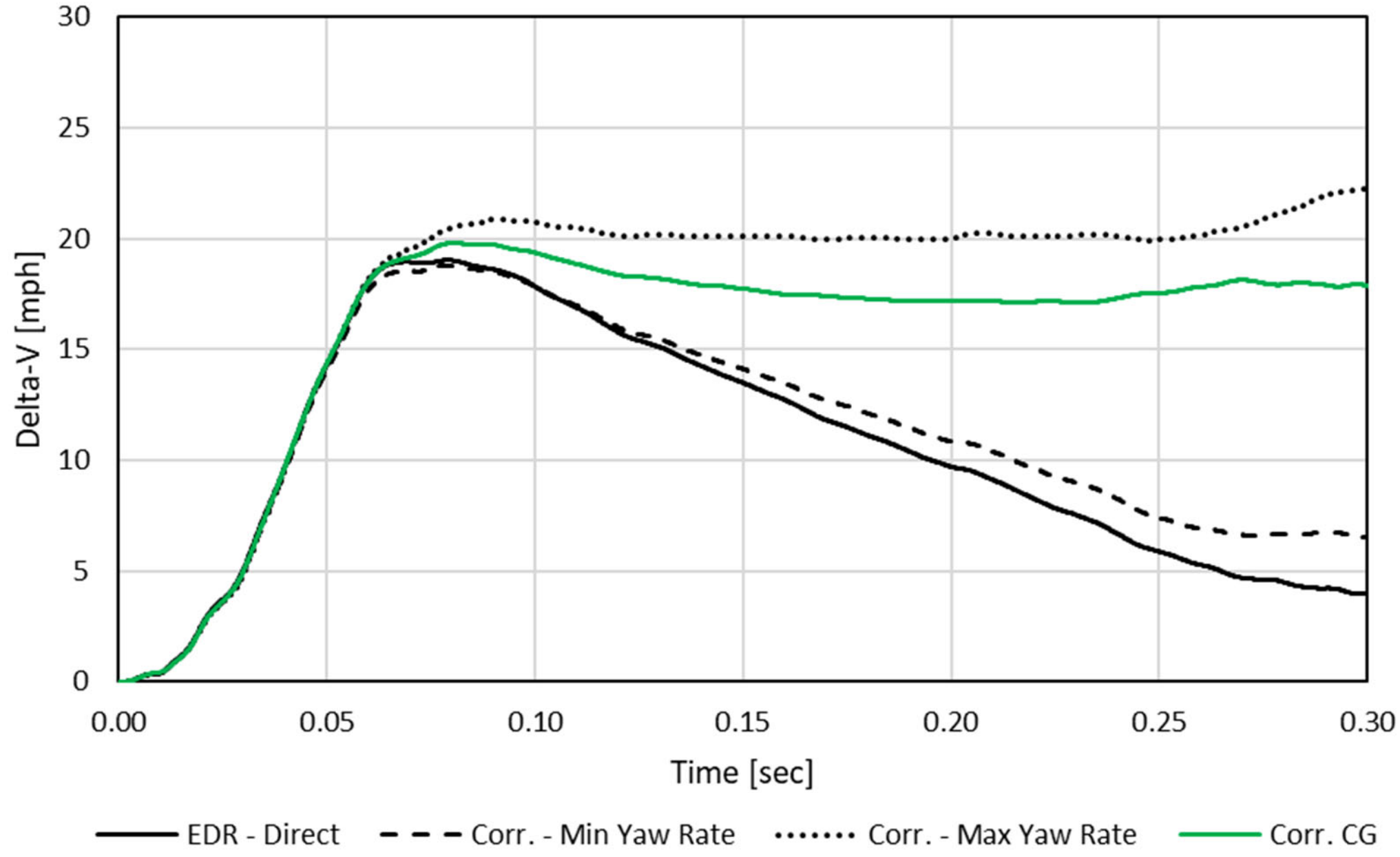

Figure 11: "Corrected" CG resultant Delta-V using equations (21) & (22), and OE EDR data

The EDR examined thus far, from a nearly 30-year-old vehicle, is located such that its position vector has both x and y components. As mentioned previously, many modern vehicles have been designed with EDR locations along, or near, the centerline of the vehicle (i.e., under the center instrument stack or under the center console). Using the data we have at hand, we can create acceleration data at two virtual EDR locations, one-foot forward of the CG along the longitudinal centerline, and one-foot aft of the CG as depicted above in the diagram of Figure 7. The resultant Delta-V corrections for these virtual EDR locations are shown respectively in Figures 12 and 13. In both cases we again see a bracketing of the actual Delta-V value by a range of about ± 2 mph. For impulses that act with moment arms aft of the CG, like the current example, the resultant CG delta-V can be expected to be greater than those directly integrated at EDR locations forward of the CG and less than those directly integrated at EDR locations aft of the CG. Conversely, for impulses that act with moment arms forward of the vehicle CG, the resultant CG Delta-V can be expected to be less than those directly integrated at locations forward of the CG and greater than those directly integrated at locations aft of the CG.

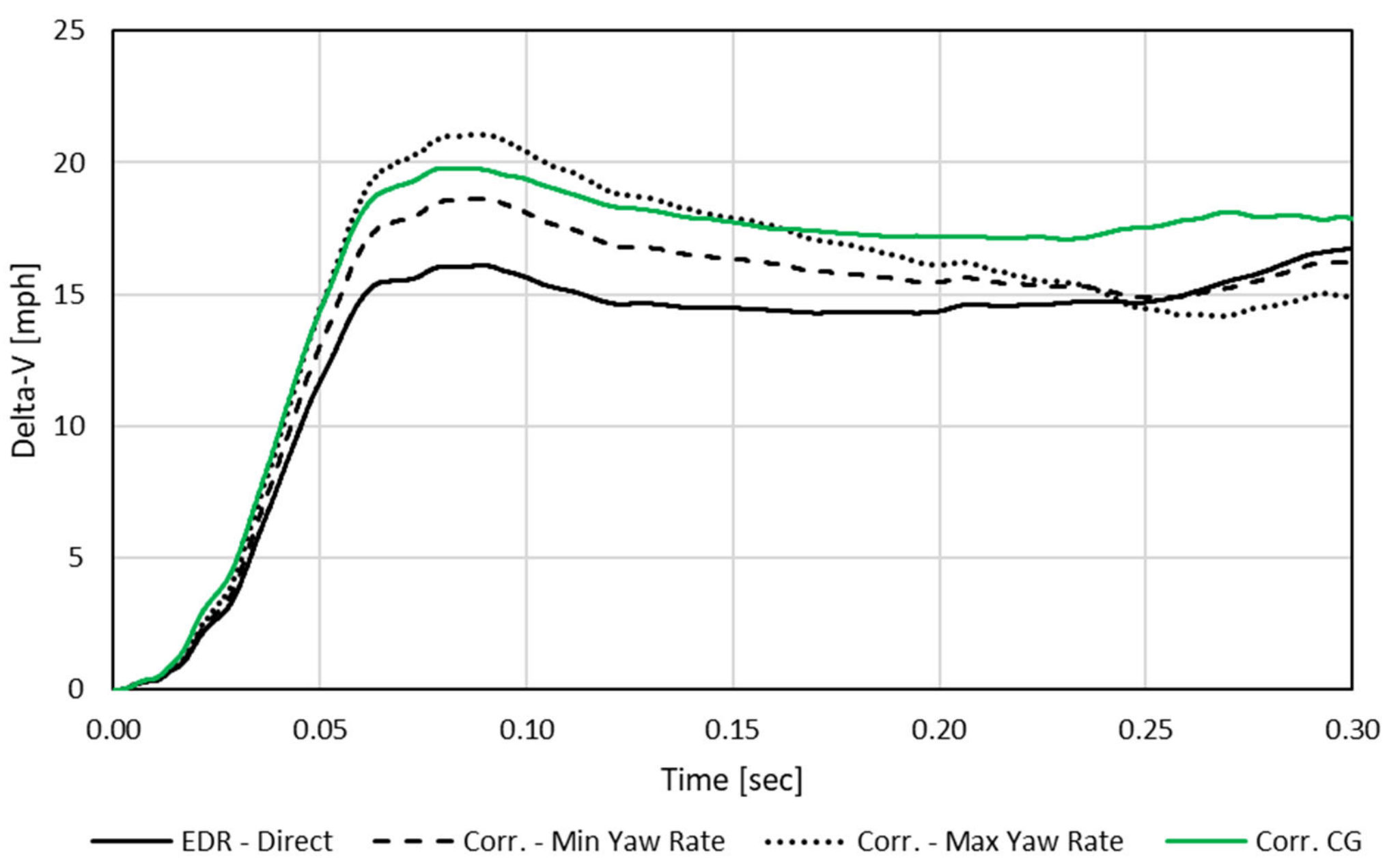

Figure 12: "Corrected" CG resultant Delta-V – virtual EDR location forward of CG

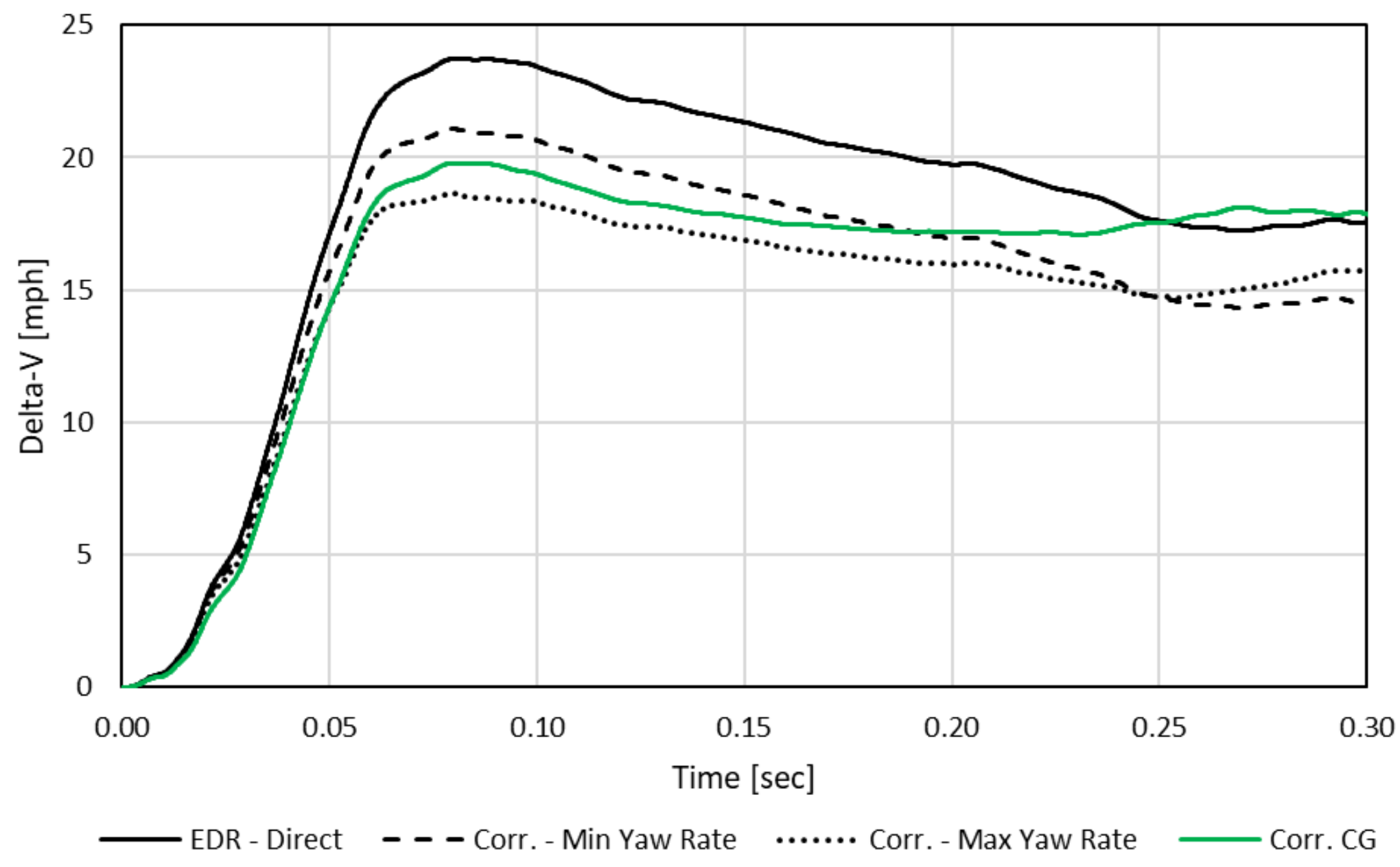

Figure 13: "Corrected" CG resultant Delta-V – virtual EDR location aft of CG

While useful results can be obtained with respect to resultant Delta-Vs, the longitudinal and lateral components of equations (17) and (20) have no practical value in the determination of a principle direction of force (PDOF) for the assessment of occupant motion within the interior of a vehicle. In the context of a vehicle undergoing substantial yaw rate and heading change, as discussed in reference [3], the notion of a PDOF becomes less meaningful for the assessment of relative occupant motion within a vehicle. In these circumstances, a free-particle analysis, first presented in a yawing-vehicle context by Wirth et al [5], is required to shed light on potential occupant motion. To conduct a free-particle analysis requires an approach like that very ably described by Smith [4]. The methods described therein require the estimation of vehicle-referenced acceleration data derived from reported EDR Delta-V data. Additionally, these methods require numerical integration of the acceleration data along with coordinate transformations that require integration of the angular velocity time-history approximation (or the formulation of a yaw angle time-history function). Rather than working through these full processes, if an analyst is working with data from an EDR located remote from the vehicle CG, and simply wants an approximation of the resultant CG Delta-V, the inaccurate directly-integrated EDR Delta-V component data may nevertheless be used to readily inform the analyst through the use a yaw rate estimate and the resultant of equations (21) and (22).